\documentstyle[11pt,paspconf,epsf]{article}
\newcommand{\eq}{\begin{equation}}
\newcommand{\en}{\end{equation}}

\begin{document}
\title{Metallicity Dependence of Starburst Spectra}
\author{Peeter Traat}
\affil{Tartu Observatory, EE2400 Tartu, Estonia}

\begin{abstract}
We have studied the impact of chemical composition on the spectrum of 
stellar population formed in the starburst. The range of metallicities 
used - $Z=0.0001\div 0.05-0.1$ covers all the compositions observed; depending 
on age, wavelength region and IMF used the UV-fluxes differ between extreme 
compositions {\it at least} tenfold, usually more. Independently of age, 
UV-fluxes shortwards of Balmer jump monotonically decline with the growth 
of metallicity. Optical and especially IR-region are also influenced, 
at $> 1$ $\mu$m fluxes typically differ $10 \div 50$ times. In IR the actually 
brightest emitting composition gets reversed with age.
\end{abstract}

Starbursts express the ununiformity and complexity of the real global star 
formation process in galaxies. They are marked by highly enhanced star 
formation activity and presence of a numerous very young massive star 
generation.   When having an usual, solar-vicinity-like bottom-heavy IMF 
they cannot be sustained by the gas resources for long.
Starbursting galaxies are of diverse sizes and range in composition from 
about solar to very metal-poor small objects; given patchiness of star 
formation process and the role of separate starbursts seems to grow 
towards smaller, lower-mass galaxies. Some very blue lowest mass nearby 
dwarfs have yet kept their matter nearly pristine (extreme cases SBS 
0335-052 and IZw18 are 41 and 50 times undersolar in metals, Thuan 
{\it et al.} 1997) and are probably undergoing the very first substantial 
star formation event in their history.

 Starburst properties get quantified with the use of spectral models.
 We have here estimated effects of chemical composition on the composite 
 spectrum 
 of formng stellar component on the basis of youngest models in our 
 multicomposite (8 compositions $Z=0.0001\div$0.1, age range 4 Myr$\div$20 Gyr,
 in $Z$=0.1 models 10 Myr$\div$13 Gyr, 6 IMF slopes, 21 
 SFR combinations, no gas-dust absorption-reemission) grid of 
 spectrophotometric evolutionary models (Traat 1996). The resulting spectral 
 flux distributions for the standard IMF are given in Fig., heavy line 
 in all panels is the reference $Z$=0.02 curve. The general trend emerging  
 is the progressive growth of absorption and decline of flux towards shorter 
 UV-wavelengths with increasing metallicity, with differences in Balmer 
 continuum slowly growing from $\sim$2 right below Balmer limit to 
 $\sim$5 near Lyman jump at 912\AA, progressing also somewhat with the 
 age of stellar generation itself. In the Lyman continuum, 
 fluxes differ more, about 10-20 times at the 
 age 4 Myr, its drop-off with time is much more rapid for compositions 
 richer in metals. At the age 10 Myr, flux output of $Z$=0.1 population 
 is lower than that of $Z$=0.0001 stars by $\sim$100 at the base of Lyman 
 jump and $\sim$6000 at the edge of He\raisebox{.6ex}{$\circ$} 504\AA break. 
 Flux emitted in He\raisebox{.6ex}{$\circ$} is weakly present in very young 
 populations with higher metallicity, as in the illustrated 4 Myr case, 
 but very rapidly fades away. Somewhat surprisingly metal-weak populations 
 are at very early ages, due  to the earlier appearance of red supergiants 
 also in the IR brighter than near-solar metallicity stars. This situation gets
 quickly reversed with aging. The possibility suspected in some 
 starbursts - the preferential formation of massive stars what lowers the IMF 
 slope $n$ - has also its own but easily accountable impact on the resulting 
 spectrum, namely, it reduces the output in IR and rises UV the more the smaller 
the $n$ value is. 
\begin{figure}[t]
\vspace*{-.71in}
\plottwo{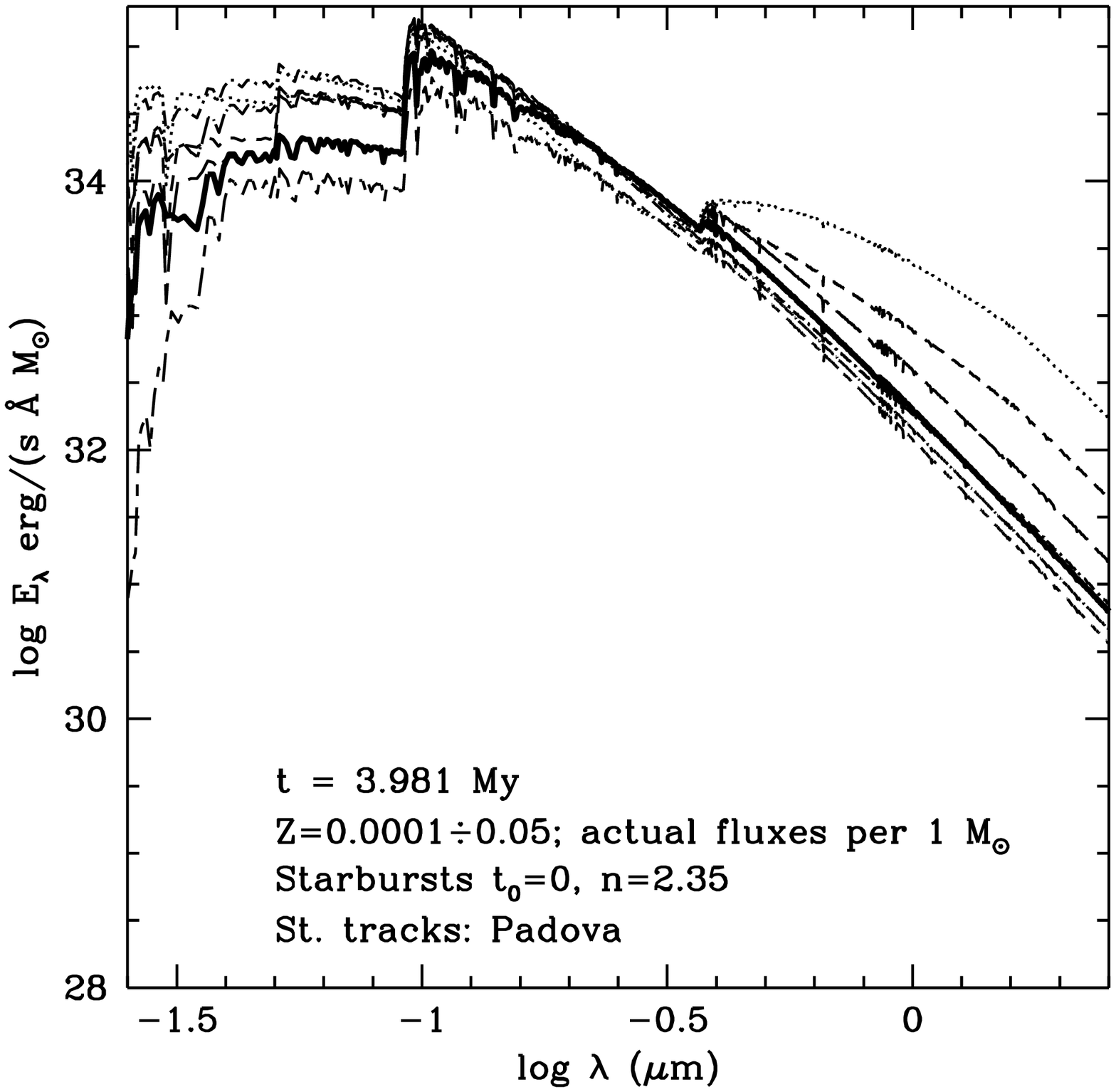}{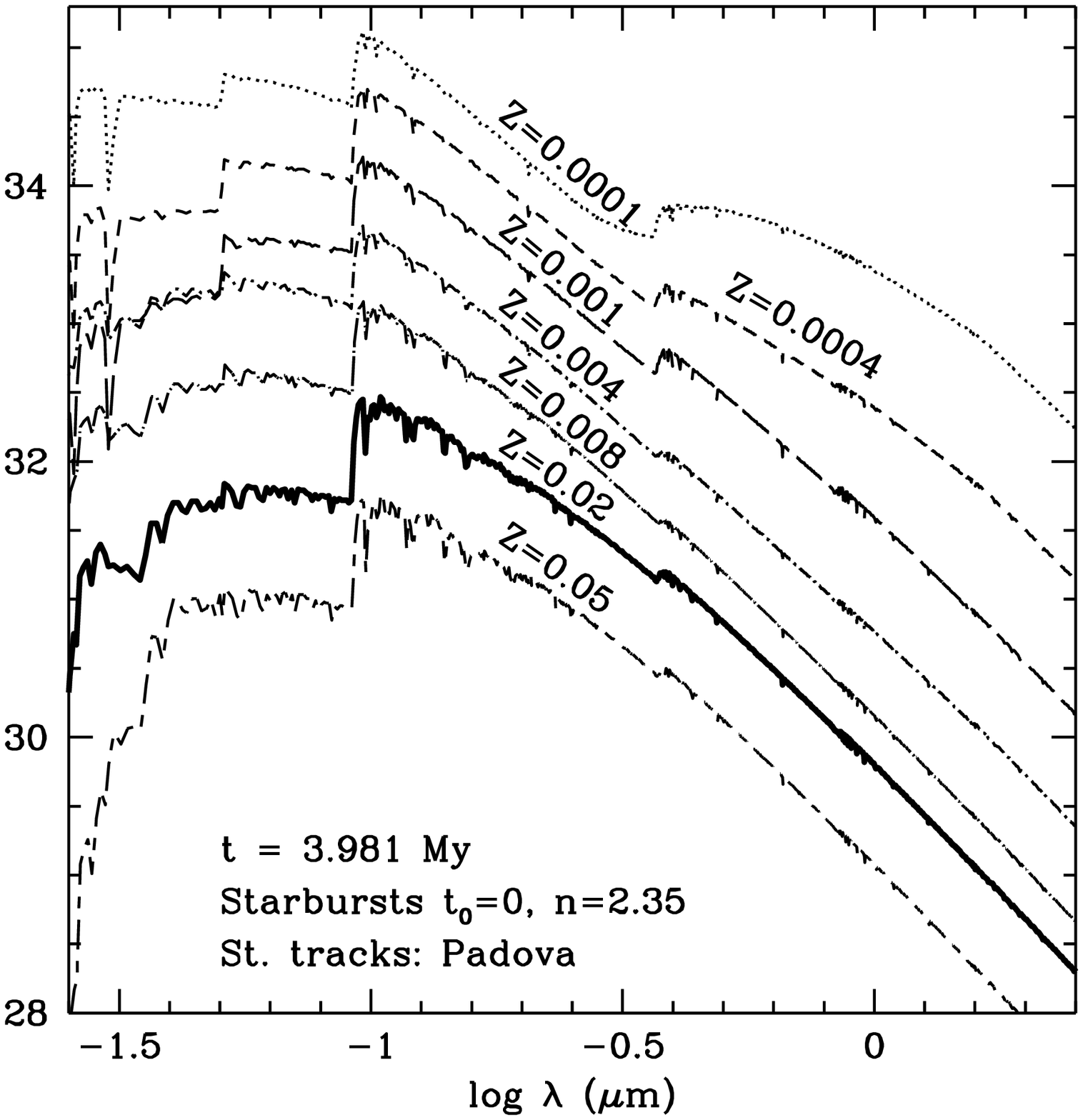}

\plottwo{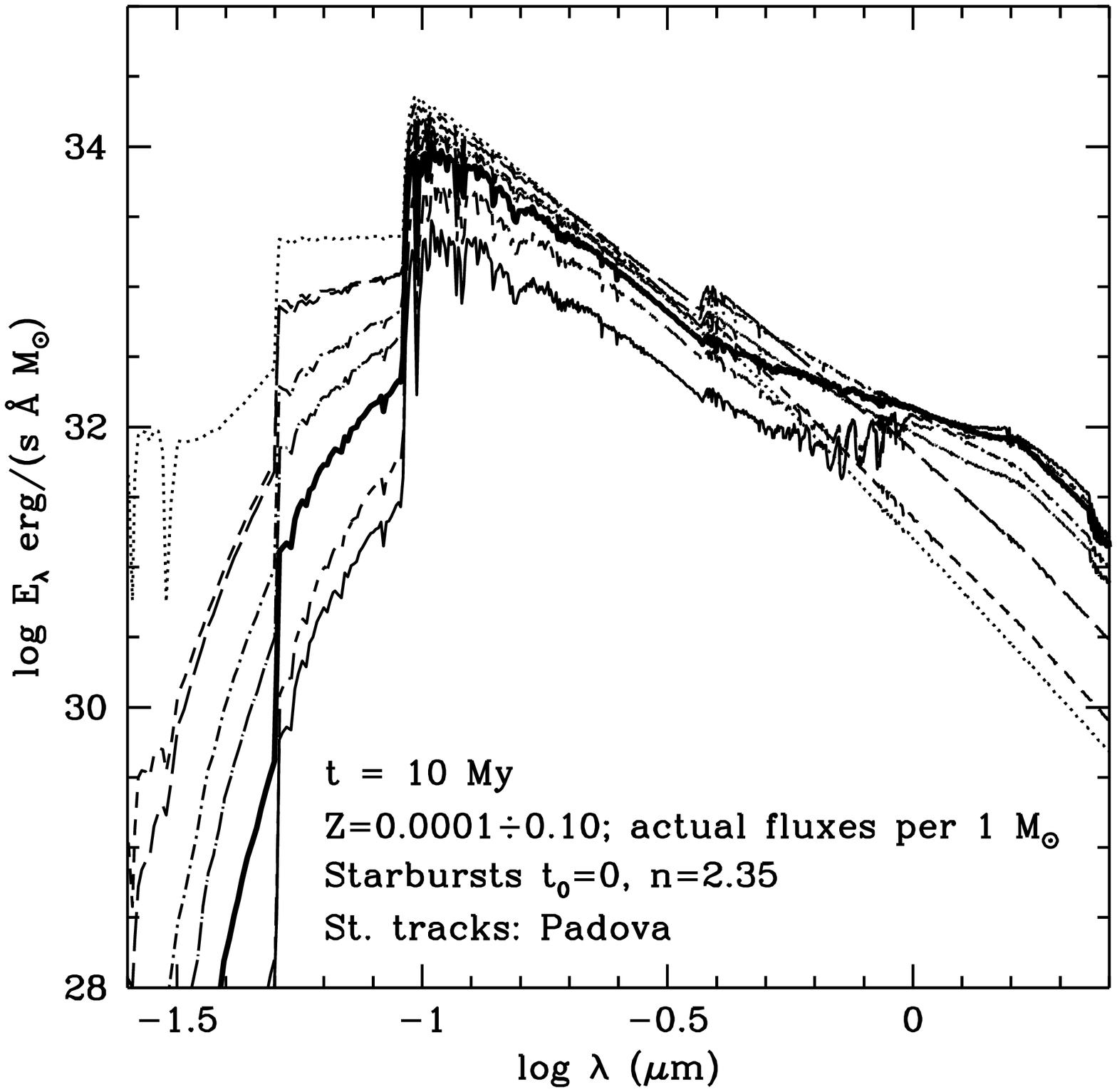}{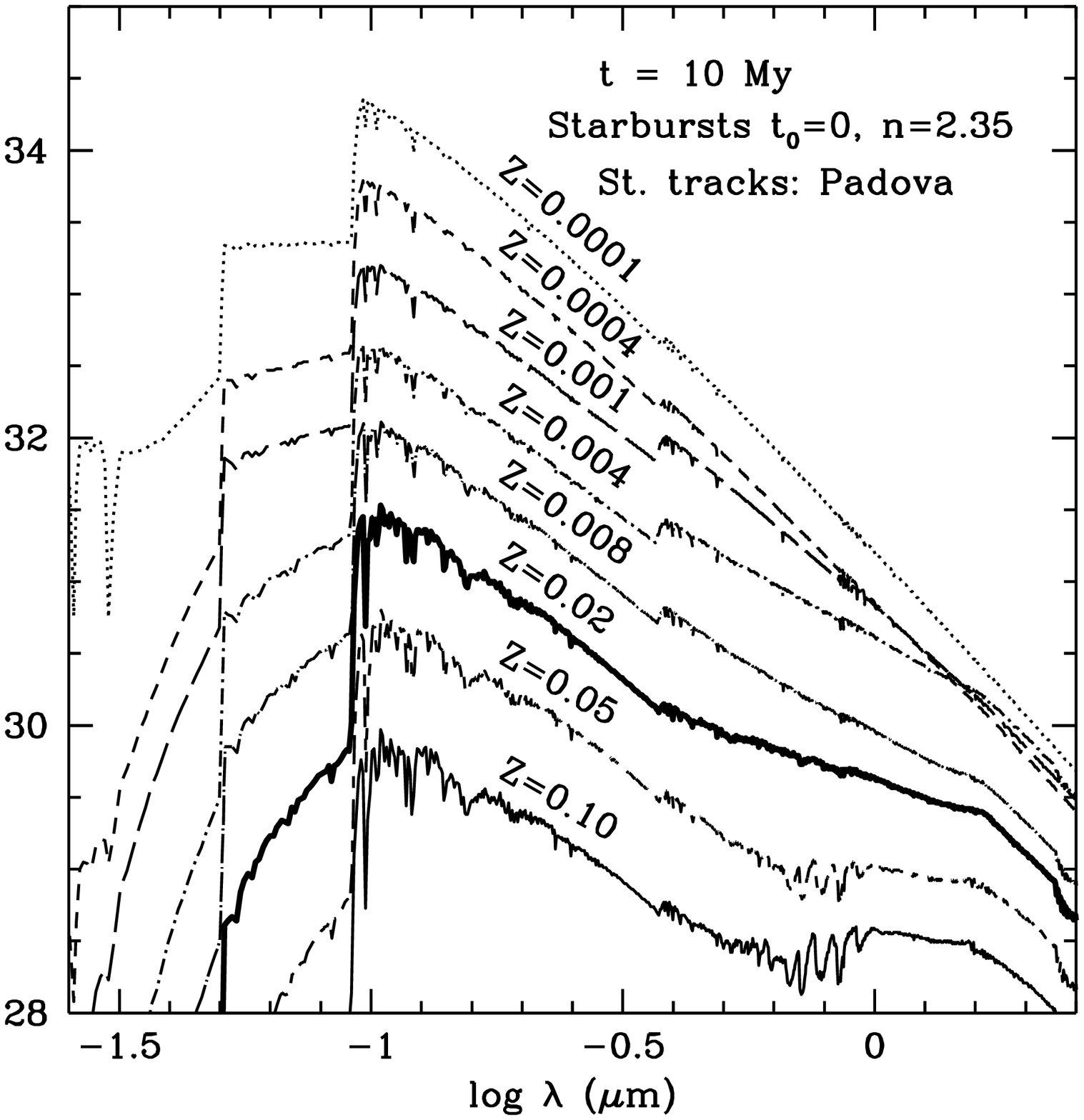}
\caption{Composition dependence of spectra of the young stellar component 
in starbursts, assumed to be formed in initial burst (SFR time-scale 
$t_0\equiv$0). Adopted IMF: power-law with the Salpeter slope ($n$=2.35); 
spectral fluxes are scaled to the unit mass in "luminous" stars 
$0.6 M_\odot \le m \le 120 M_\odot$. Original spectra for different 
compositions are depicted on the {\it left-hand panels}. They are redrawn 
for clarity on {\it right-hand panels} with successive downward shifts 
-0.5 dex to the previous one, with $Z$=0.0001 curve retaining its true 
location and $Z$=0.1 accumulated the maximum shift -3.5 dex.}
\end{figure}

\end{document}